\documentclass[journal=nalefd,manuscript=letter]{achemso}
\usepackage{wrapfig}
\usepackage{amsmath}
\usepackage{xcolor}
\usepackage{braket}

\author{Arianne Brooks}
\author{Xiao-Liu Chu}
\email{xchu@ic.ac.uk}
\affiliation[Niels Bohr Institute]
{Center for Hybrid Quantum Networks (Hy-Q), Niels Bohr Institute,
University of Copenhagen, Blegdamsvej 17, DK-2100 Copenhagen, Denmark}
\alsoaffiliation[Imperial College London]
{Present address: MRC London Institute of Medical Sciences, Du Cane road, London, W12 0NN, United Kingdom}
\author{Zhe Liu}
\affiliation[Niels Bohr Institute]
{Center for Hybrid Quantum Networks (Hy-Q), Niels Bohr Institute,
University of Copenhagen, Blegdamsvej 17, DK-2100 Copenhagen, Denmark}
\author{R\"udiger Schott}
\author{Arne Ludwig}
\author{Andreas D. Wieck}
\affiliation[Bochum]
{Lehrstuhl für Angewandte Festkörperphysik, Ruhr-Universität Bochum, Universitätsstrasse 150, D-44780 Bochum, Germany}
\author{Leonardo Midolo}
\author{Peter Lodahl}
\affiliation[Niels Bohr Institute]
{Center for Hybrid Quantum Networks (Hy-Q), Niels Bohr Institute,
University of Copenhagen, Blegdamsvej 17, DK-2100 Copenhagen, Denmark}
\author{Nir Rotenberg}
\affiliation[Niels Bohr Institute]
{Center for Hybrid Quantum Networks (Hy-Q), Niels Bohr Institute,
University of Copenhagen, Blegdamsvej 17, DK-2100 Copenhagen, Denmark}
\alsoaffiliation[Queens University]
{Present address: Department of Physics, Engineering Physics \& Astronomy, 64 Bader Lane, Queen’s University, Kingston, Ontario, Canada K7L 3N6
 }

\title{An integrated whispering-gallery-mode resonator for solid-state coherent quantum photonics}

\abbreviations{IR,NMR,UV}

\begin{document}


\begin{abstract}
Tailored photonic cavities allow enhancing light-matter interaction ultimately to create a fully coherent quantum interface. Here, we report on an integrated microdisk cavity containing self-assembled quantum dots to coherently route photons between different access waveguides. We measure a Purcell factor of $F_{exp}=6.9\pm0.9$ for a cavity quality factor of about 10,000, allowing us to observe clear signatures of coherent scattering of photons by the quantum dots. We show how this integrated system can coherently re-route photons between the drop and bus ports, and how this routing is controlled by detuning the quantum dot and resonator, or through the strength of the excitation beam, where a critical photon number less than one photon per lifetime is required. We discuss the strengths and limitations of this approach, focusing on how the coherent scattering and single-photon nonlinearity can be used to increase the efficiency of quantum devices such as routers or Bell-state analyzers.
\end{abstract}

\section{Introduction}
Photonic resonators enhance light-matter interactions, and have played a crucial role in quantum optical experiments over the past several decades. Resonators such as photonic crystal cavities \cite{Akahane:03} or whispering gallery mode resonators \cite{Armani:03} have been fabricated on photonic chips, leading to pioneering demonstrations of strong light-matter coupling of single atoms \cite{Aoki:06} and quantum dots (QDs) \cite{Hennessy:07,Loo:10}, or an increase in the coherent interaction between photons and single organic molecules \cite{Wang:19}. Whispering gallery mode resonators also support chiral quantum interactions \cite{Lodahl:17,Martincano:19}, where photons are emitted or scattered unidirectionally, enabling non-reciprocal photonic elements constructed with single emitters such as optical circulators \cite{Scheucher:16}, isolators \cite{Sayrin:15} and atom-photon SWAP gates \cite{Bechler:18}. 

Here, we create an integrated photonic circuit consisting of a microdisk resonator with embedded self-assembled QDs, access waveguides and grating couplers, as shown in Fig.~\ref{setup}a. The enhancement provided by the resonator lessens the effect of decoherence mechanisms\cite{Pedersen:20}, most notably spectral diffusion, enabling the observation of coherent scattering of photons from the QD and leading to a coherent switching of photons between the bus and drop ports. This stands in contrast to earlier demonstrations where QDs embedded in a photonic crystal cavity could modulate the transmission across a single channel coupled to the cavity \cite{Englund:12, Sun:18}. Cryogenic spectroscopy and time-resolved measurements in conjunction with quantum optical theory allow us to quantify the effect of the resonators on the QD, and to explore the response of the QD-resonator detuning and excitation strengths on the photon routing.

\section{Integrated microdisk resonators}

\begin{figure}
    \includegraphics[trim={0 0 0 0},clip, width=0.45\textwidth]{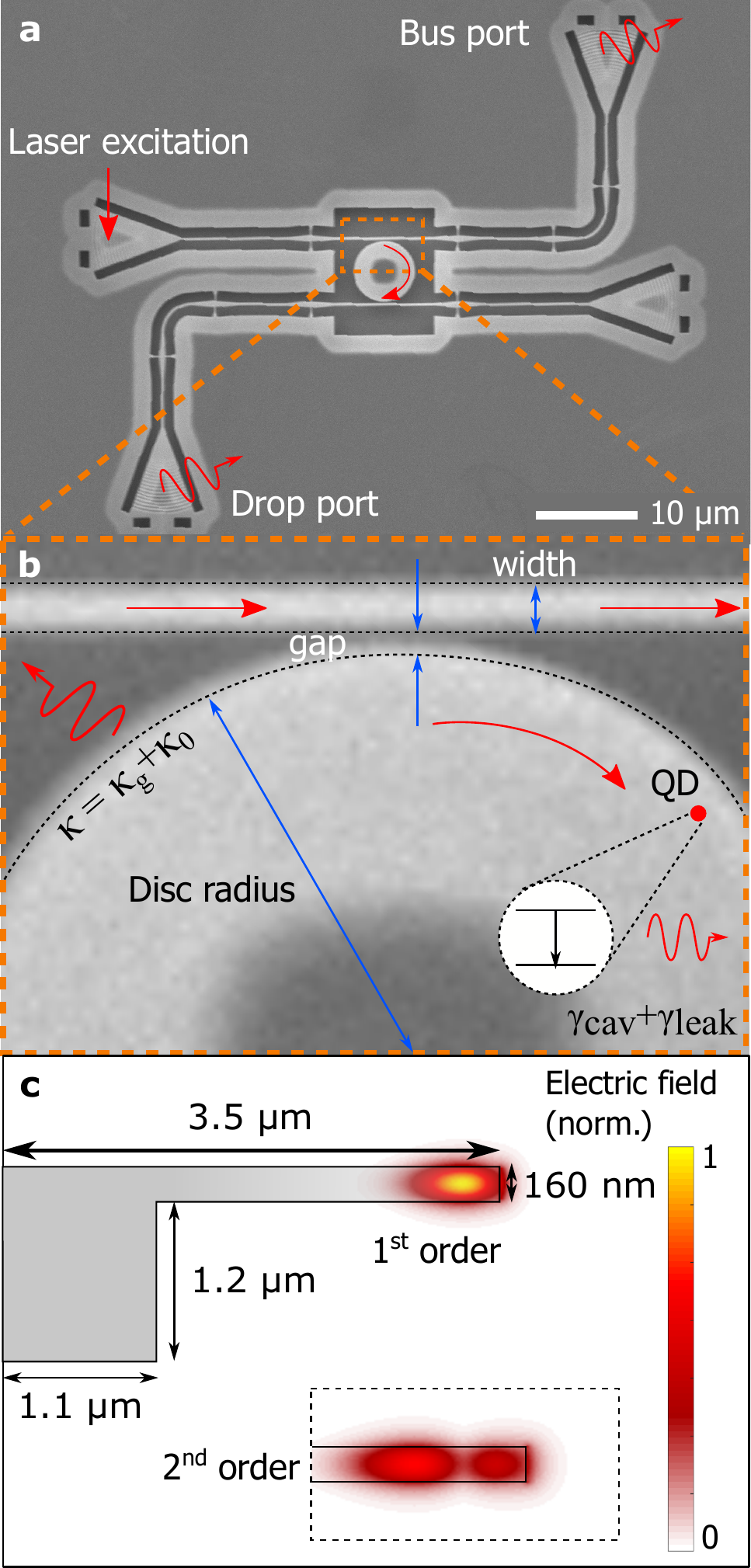}
   \caption{a) Scanning electron microscope image of the integrated disk cavity, showing the excitation, bus and drop ports. b) Zoom-in of the disk resonator and access waveguide, with a schematic of the QD position as indicated. The QD acts as a two-level system that emits into the cavity with rate $\gamma_{\mathrm{cav}}$ and into free-space with rate $\gamma_{\mathrm{leak}}$, while the cavity loss rate $\kappa$ arises as a combination of scattering into free-space with rate $\kappa_\mathrm{0}$ and coupling into the access waveguides with rate $\kappa_\mathrm{g}$. The QD couples to the optical modes of the resonator, which are calculated using finite element methods and shown in (c).}
  \label{setup}
\end{figure}

We fabricate GaAs disk-shaped cavities that support whispering gallery modes that are optically addressed via evanescently coupled single mode waveguides, as shown in Figure \ref{setup}a). In the present sample no electrical contacts were implemented, which otherwise have been shown to efficiently overcome QD broadening due to electrostatic charge fluctuations \cite{Thyrrestrup:18}. However, electrically contacted samples may increase absorption losses and increases fabrication complexity \cite{Wang:21}, such that an alternative strategy using Purcell enhancement to reduce the influence of noise processes is a favorable approach. The high intrinsic quantum efficiency of QDs means that any non-radiative processes can be neglected (c.f. Supplementary Information) \cite{Wang:11}. Consequently, we need only consider radiative decay, which occurs with rates $\gamma_{\mathrm{cav}}$ and $\gamma_{\mathrm{leak}}$ into the resonator modes and free-space, respectively, as depicted in Figure \ref{setup}b).

The disk resonators are fabricated with a 3.5~$\mu$m radius, chosen because a $\approx 1.1$~$\mu$m support pillar remains after the disk and waveguides are under-etched (dark region in Figures~\ref{setup}a and b), and because finite element simulations (COMSOL Multiphysics) reveal negligible bending losses. In fact, for the first two radial modes, as shown in Figure~\ref{setup}c, we find intrinsic quality factors (Q-factor) limited only by the computational accuracy $\left(Q_{\mathrm{theory}} \approx 10^{13} \right)$. This value is well above typically reported values of $Q = 10^5$ for QD-based GaAs resonators \cite{Gayral:01,Michael:07,Baker:11} limited by surface roughness and gap state related surface absorption \cite{Najer:21}. However, these effects can be decreased by employing surface passivation techniques, resulting in ultrahigh Q-factor resonators $\left(Q\geq10^{6}\right)$ \cite{Guha:17}. In our case, a further reduction due to coupling between the resonator and the access waveguides is expected. From the field distributions, we calculate the effective mode volumes\cite{Sauvan:13} of the first and second radial modes $V_{\mathrm{eff}}^{\left(1\right)} \approx 18(\lambda/n)^3$ and $V_{\mathrm{eff}}^{\left(2\right)} \approx 22(\lambda/n)^3$ (c.f. Supplementary Information).

To characterize the integrated photonic resonator, we use an optimized grating coupler \cite{Zhou:18} to launch light from a tunable continuous-wave laser through the access waveguides and into the disk. The access waveguide is single mode at the 940 nm emission wavelength of the QDs, and is tapered to a width of 220 nm in the vicinity of the resonator to improve coupling to the cavity mode. Additionally, in a series of different structures, the gap between the disk and waveguides is varied between 40 and 160 nm, in steps of 30 nm, to determine the critical coupling geometry.

\begin{figure*}[t]
  \includegraphics[trim={0 0 0cm 0cm},clip, width=1\textwidth]{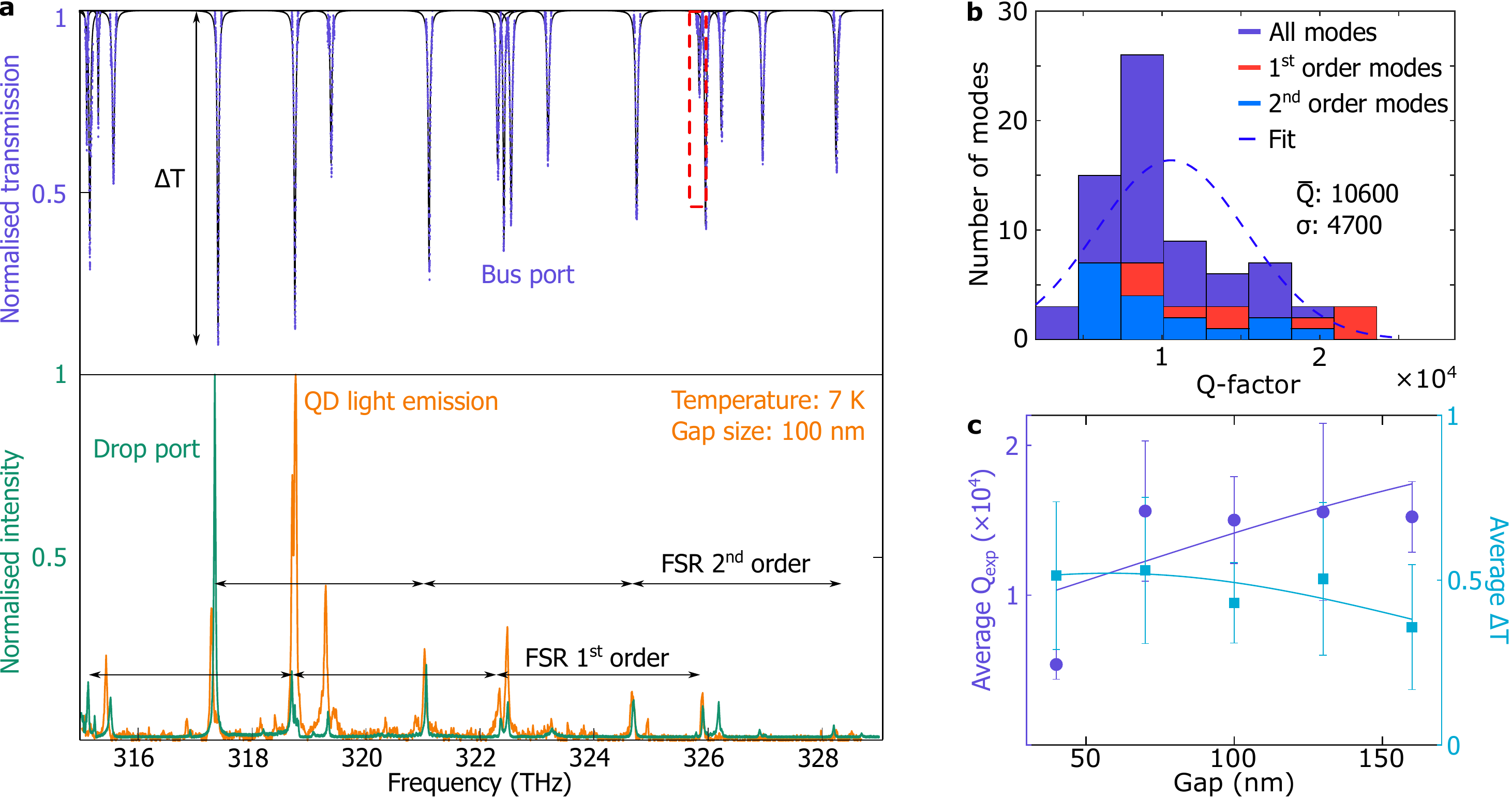}
  \caption{a) Exemplary bus (purple) and drop (green) port transmission intensity as a function of frequency, measured by scanning the laser and collecting light from respective port. Signatures of the optical resonances are clearly visible in both, and their separations agree well with the calculated free spectral range of the disk. For comparison, the emission spectrum for QDs excited non-resonantly and measured through the drop port is also presented (orange), showing strong emission enhancement when the QDs are on resonance. The first-order resonance that is coupled to a QD is marked with the red dashed box. b) Histogram of first-order, second-order and all mode Q-factors extracted from the bus port data for all structures, results in a mean Q in excess of 10,000. c) Dependence of $\bar{Q}$ (left axis) and average $\Delta T$ on the structure gap width. Error bars represent the statistical variance, while solid lines are theoretical fits from Eq. \ref{criticalT}.
  }
  \label{largescan}
\end{figure*}

Working at cryogenic temperatures, we scan the excitation laser frequency over a 13 THz bandwidth and record the outcoupled intensity transmitted through both the bus and drop ports, as shown in Figure~\ref{setup}a. Exemplary spectra are shown in Figure~\ref{largescan}a, here for a gap size of 100 nm. In the bus port spectrum, which has been normalized to a highly dispersive background (all dips are shown; see Supplementary Information for raw data), we observe sharp dips at the WGM frequencies where the disk couples light from one access waveguide to another. Since the different resonator modes couple to the access waveguides with different efficiencies, the depths of the dips $\Delta T$ vary. As expected, the dips in the bus port spectrum are well correlated  with peaks in the drop port spectrum, and the different resonance orders can be determined by the measured free spectral range. Furthermore, the QDs in the cavity are excited non-resonantly using a Ti:Sapphire laser at 810 nm (Tsunami) and the emission collected on a spectrometer. Note the strong emission enhancement when the QDs are on resonance with a cavity mode.

These measurements are repeated for all structures, fitting each cavity resonance with a Lorentzian function to determine its width $\kappa$, allowing us to deduce the loaded Q-factor $Q_{\mathrm{exp}}=\omega_\mathrm{c}/\kappa$, where $\omega_\mathrm{c}$ is the central resonance frequency. A collection of $Q_{\mathrm{exp}}$ values are shown in a histogram in Figure \ref{largescan}b), where a mean of 10600 $\pm$ 4700 is obtained and the largest average $\bar{Q}_{\mathrm{exp}}$ is measured for 1st order modes ( $\bar{Q}^{\left(1\right)} = $13600 $\pm$ 5400 vs $\bar{Q}^{\left(2\right)} = $9300 $\pm$ 4900). 

To determine the optimal, critically coupled configuration, we consider the gap-width dependence of both $\bar{Q}$ and change in transmission $\Delta T$, cf. Figure~\ref{largescan}c. Qualitatively, as the gap size increases, leading to a weaker coupling between the resonator and access waveguides, the signature of the coupling $\Delta T$ decreases with a corresponding increase in $\bar{Q}$. This trend agrees well with the theoretical prediction (solid curves) for the loaded ring resonator,\cite{Ding:10}
\begin{align}
    \Delta T &= 1-\left[T_{cc}+(1-T_{cc})\left(\frac{1-\kappa_g}{1+\kappa_g}\right)^2\right],\\
    \frac{1}{Q_{exp}} &= \frac{1}{Q_{int}}\left(1+\kappa_g\right),
    \label{criticalT}
\end{align}
where $\kappa_\mathrm{g} = \kappa_{\mathrm{g0}}e^{-\xi g}$ characterizes the coupling rate between the cavity and access waveguides, $\xi$ is the characteristic length constant, and $g$ is the gap size. In these equations, $T_{\mathrm{cc}}$ is the transmission at critical coupling, while $Q_{\mathrm{int}}$ is the intrinsic Q-factor of the resonator (i.e. in the absence of the access waveguides). From modelling the data, we find $\bar{Q}_{\mathrm{int}} = (2.3 \pm 0.1)\times10^4$ and a critical coupling gap size of $\sim 64 \pm 10$~nm, well within the reach of modern nanofabrication techniques. 

\section{Resonant scattering from a quantum dot}
\begin{figure*}[ht!]
  \includegraphics[trim={0 0cm 0 0},clip, width=1\textwidth]{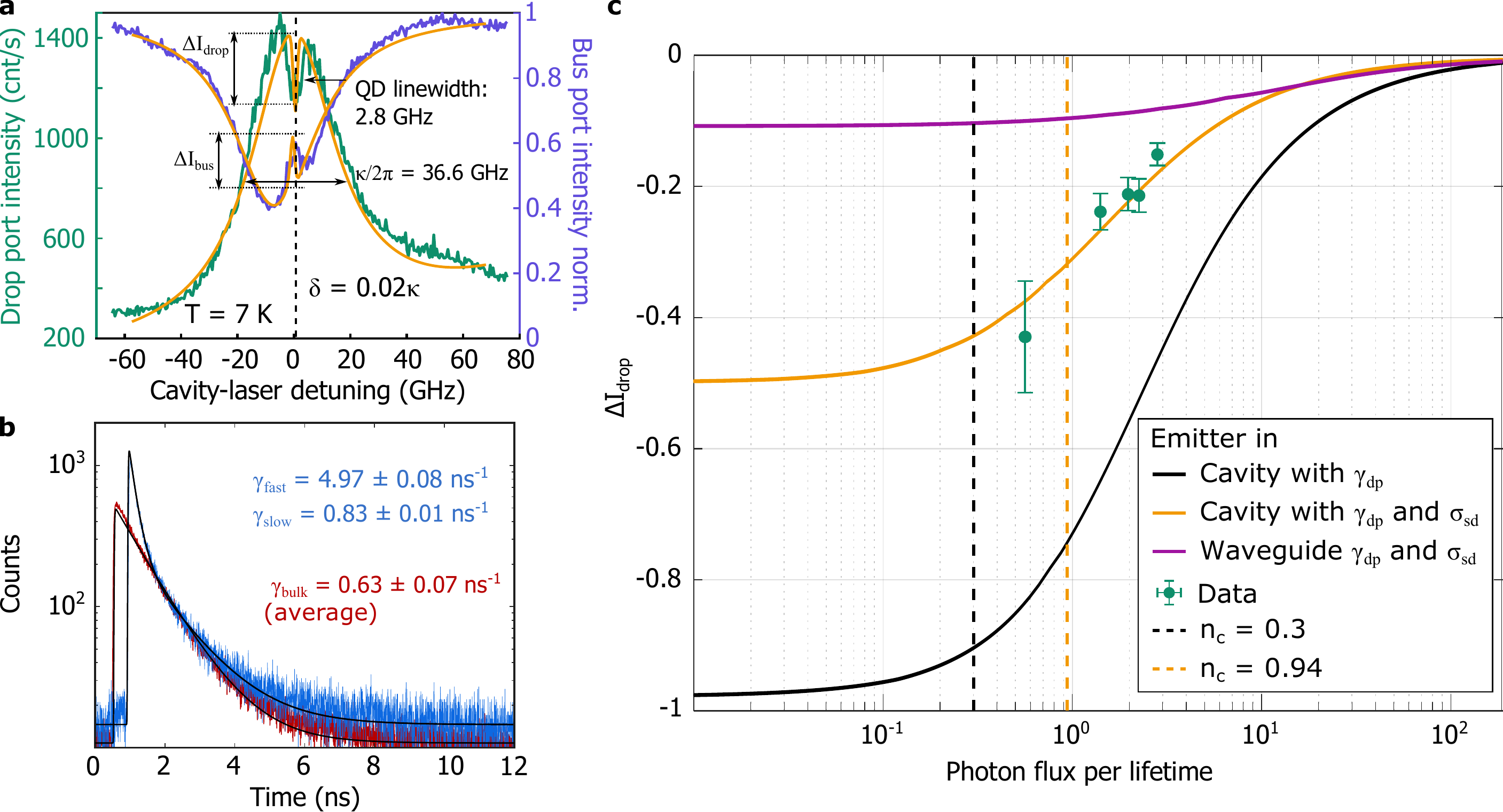}
  \caption{a) Low-power, frequency-dependent intensity as a function of cavity-laser detuning recorded both from the drop port (green) and bus port (purple), taken at 7K when the QD and cavity are nearly on resonance. Fits to theory (orange curves, Eq. \ref{eq:T_conv}) enable extracting parameters such as the QD and cavity linewidths, here 2.8 GHz and 36.6 GHz, respectively. b) Lifetime measurement for the QD on resonance with the cavity (blue) and in a bulk sample (red) with corresponding fits. c) Power-dependent change in transmission of the QD of (a), showing a clear decrease in extinction for higher incident photon fluxes. Also shown is the theoretical transmission (see main text) for the QD-cavity system accounting only for dephasing (black) and also for spectral diffusion (orange). For the latter, a critical photon number of 0.9 photons per lifetime is found (orange dashed line). For comparison, the predicted saturation curve for a QD in a waveguide (i.e. with no Purcell enhancement) is shown (purple).
  }
  \label{7K}
\end{figure*}
We now turn to the QDs embedded within the disks and study how the resonators alter the light-matter interaction. Figure~\ref{7K}a) shows typical drop (left axis, green) and bus (right axis, purple) intensity as a function of cavity-laser detuning, taken on a sample with a 100 nm gap at 7 K and at 5 $\mu$W excitation power. In this figure, we see a clear signature of the coherent interaction between photons and the QD (highlighted by dashed lines in Fig.~\ref{7K}a), resulting in a re-routing of the photons between the bus and drop ports, at a QD-cavity detuning of $\delta = 0.02 \kappa$. In the bus port, we observe a clear extinction by the QD of the transmission that is indicative of interference between the photons scattered by the QD and the incoming probe field \cite{Turschmann:17, Thyrrestrup:18}. Similarly, we observe a peak in the bus port intensity at the same location, as additional photons are scattered into this channel by the QD.

To accurately model the frequency response presented in Figure~\ref{7K}a we first require knowledge of the emitter decay rate, which is the sum of decay rates into free-space $\gamma_{\mathrm{leak}}$ and the cavity mode $\gamma_{\mathrm{cav}}$. We therefore measure the QD lifetime in both bulk GaAs and when coupled to the microdisk, presenting exemplary results in Figure~\ref{7K}b. Bulk (red data) measurements are well-fitted by a single-exponential decay with an average value of $\gamma_{\mathrm{bulk}}$  = (0.63 $\pm$ 0.07) ns$^{-1}$, corresponding to the natural linewidth of $\gamma_{\mathrm{bulk}}/2\pi$  = (0.1 $\pm$ 0.01) GHz. In contrast, a double-exponential is needed to fit the cavity enhanced lifetime measurement (blue data), which we attribute to the different coupling of the two, orthogonally polarized QD transition dipoles to the cavity. Here, one dipole is well coupled to the cavity and hence has a fast decay rate $\gamma_{\mathrm{fast}} = \gamma_{\mathrm{cav}} +\gamma_{\mathrm{leak}} = (4.97 \pm 0.08$) ns$^{-1}$ ((0.79 $\pm$ 0.01) GHz linewidth), while the other is weakly coupled with a decay rate $\gamma_{\mathrm{slow}} = (0.83 \pm 0.01$) ns$^{-1}$ ((0.31 $\pm$ 0.002) GHz linewidth). By comparing the decay rate of the well-coupled transition $\gamma_{\mathrm{fast}}$ to that of bulk $\gamma_{\mathrm{bulk}}$, we find a lifetime enhancement of 7.9 due to the cavity. While it is likely that embedding the QD in the microdisk suppresses emission into free-space, relative to an emitter in the bulk, in what follows we assume that $\gamma_{\mathrm{leak}} \approx \gamma_{\mathrm{bulk}}$ as is done in literature\cite{Srinivasan:07}, which means that we extract lower-bounds on the Purcell factor and the coupling efficiency of our system. Finally, we take the pure-dephasing rate for the QD embedded in the microdisks and at temperatures ranging from 6 - 12 K to be $\gamma_{\mathrm{dp}}/2\pi = 0.01$~GHz, as reported in literature \cite{Tighineanu:18}.

Having determined $\gamma_{\mathrm{cav}}$, $\gamma_{\mathrm{leak}}$ and $\gamma_{\mathrm{dp}}$, we repeat the spectral measurements such as those presented in Figure~\ref{7K}a, increasing the excitation laser. For the drop port (green data), for example, the transmitted intensity is $T_\mathrm{drop}=\eta \left|t_\mathrm{drop}\right|^2$, where $\eta$ accounts for the incidence photon flux and the cavity-mediated coupling efficiency between the bus and drop port waveguides. An analytic form of the transmission coefficient, including coherent scattering from the QD, is known to be \cite{Garnier:07,Wang:17}
\small
\begin{align}
  &t_\mathrm{drop} =t_0 [ -1 \nonumber \\
  &+\frac{f}{(1+S)(f+(1+\frac{2i\Delta\omega}{(\gamma_\mathrm{leak}+2\gamma_\mathrm{dp})})(1+i\frac{\Delta\omega+\delta}{(\kappa/2)}))}], 
  \label{transmissioncoeff}
\end{align}
\normalsize
where $\Delta\omega = \omega_\mathrm{laser}-\omega_\mathrm{QD}$ is the laser detuning to the QD resonance, $f = \gamma_{\mathrm{cav}} / \left(\gamma_{\mathrm{leak}}+2\gamma_{\mathrm{dp}}\right)$, $t_0 = 1 / \left[1 + i\left(\Delta\omega+\delta\right) / \left(\kappa/2\right)\right]$ and $S$ is the saturation parameter that accounts for the incident power (see Supplementary Information for relationship of $S$ to input power and photon number per lifetime, and the corresponding $t_\mathrm{bus}$). Spectral diffusion in the system results in `wandering' of the QD resonance, which can be modelled by a convolution of the transmission with a Gaussian with linewidth $\sigma_{\mathrm{sd}}$:\cite{lejeannic:21}
\begin{equation}
    T_\mathrm{drop,conv} = |t_\mathrm{drop}|^2 * P(\sigma_\mathrm{sd}),
    \label{eq:T_conv}
\end{equation}
where 
\begin{equation}
    P(\sigma_\mathrm{sd})=\frac{1}{\sqrt{2\pi}\sigma_\mathrm{sd}}\mathrm{exp}\bigg(-\frac{1}{2}\big(\frac{\Delta\omega-\delta}{\sigma_\mathrm{sd}}\big)^2\bigg).
\end{equation}
As can be seen in Figure~\ref{7K}a, the frequency response is well reproduced by the theory.

In practise, bus and drop port frequency-resolved data at different excitation powers are simultaneously fit with $\delta$, $\omega_{QD}$, $\kappa$, $S$ and $\sigma_{\mathrm{sd}}$ as free parameters, noting that $\sigma_{\mathrm{sd}}$ is temperature dependent (see Figure 6 in the Supplementary Information). For the 5~$\mu$W data presented in Figure~\ref{7K}a, we find S = 1.5 $\pm$ 0.2 (corresponding to 1.4 $\pm$ 0.2 photons per lifetime), a QD-cavity detuning of $\delta=0.02\kappa$ where $\kappa / 2 \pi = 36.6\pm 2$ GHz and spectral diffusion of $\sigma_{\mathrm{sd}}/2\pi= 0.6 \pm 0.1$ GHz. We also find a coherent extinction of photons in the drop port ($\Delta I_{\mathrm{drop}}$) of (-24 $\pm$ 4)$\%$, as those photons are re-routed back into the bus port ($\Delta I_{\mathrm{bus}}$) by the QD. The measured ratio of $\sigma_{\mathrm{sd}}/\gamma_{\mathrm{cav}} \approx$0.87 is a factor of 4 better than what has been achieved in slow-light photonic crystals with QDs that are not electrically contacted \cite{Javadi:15}, where a peak extinction of $8\%$ was observed.

The routing can be controlled either through the QD-cavity detuning or by varying the intensity of the incident photon stream.  We first demonstrate the latter, presenting the fraction of photons re-routed from the drop to bus ports as a function of the incident photon flux per lifetime in Figure~\ref{7K}c. Here, the extinction measurements (symbols) are compared with three theoretical predictions (using Eq.~\ref{transmissioncoeff} above and the fitted parameters): the QD-resonator with broadening due to pure dephasing and spectral diffusion (orange curve), with pure dephasing only (as for an electrically contacted sample, black curve), or a QD in a waveguide (i.e. no emission enhancement, purple curve). The measurements are well reproduced when both pure dephasing and spectral diffusion are accounted for, and for this detuning $\left(\delta = 0.02\kappa\right)$ a critical number of photons per lifetime of $n_c = 0.94\pm 0.2$ and maximum extinction of $\Delta I_{\mathrm{drop, max}} = (-53 \pm 4)\%$ in the limit of low power ($S=0$) were found. For comparison, the maximum extinction realizable for a QD without Purcell enhancement is (-$15 \pm 4$) \%. For an electrically contacted QD system, where $\sigma_{sd}=0$, the critical photon number is expected to decrease to $n_c = 0.3 \pm 0.02$, with a maximum $\Delta I_{\mathrm{drop, max}} = (-98 \pm 2 )\%$ being achievable (black curve) (see Supplementary Information). These results benchmarks the conditions for coherent routing of photons in between the bus and drop ports at the single-photon level.

\begin{figure}[t!]
  \begin{center}
    \includegraphics[trim={0 0 0 0},clip, width=0.45\textwidth]{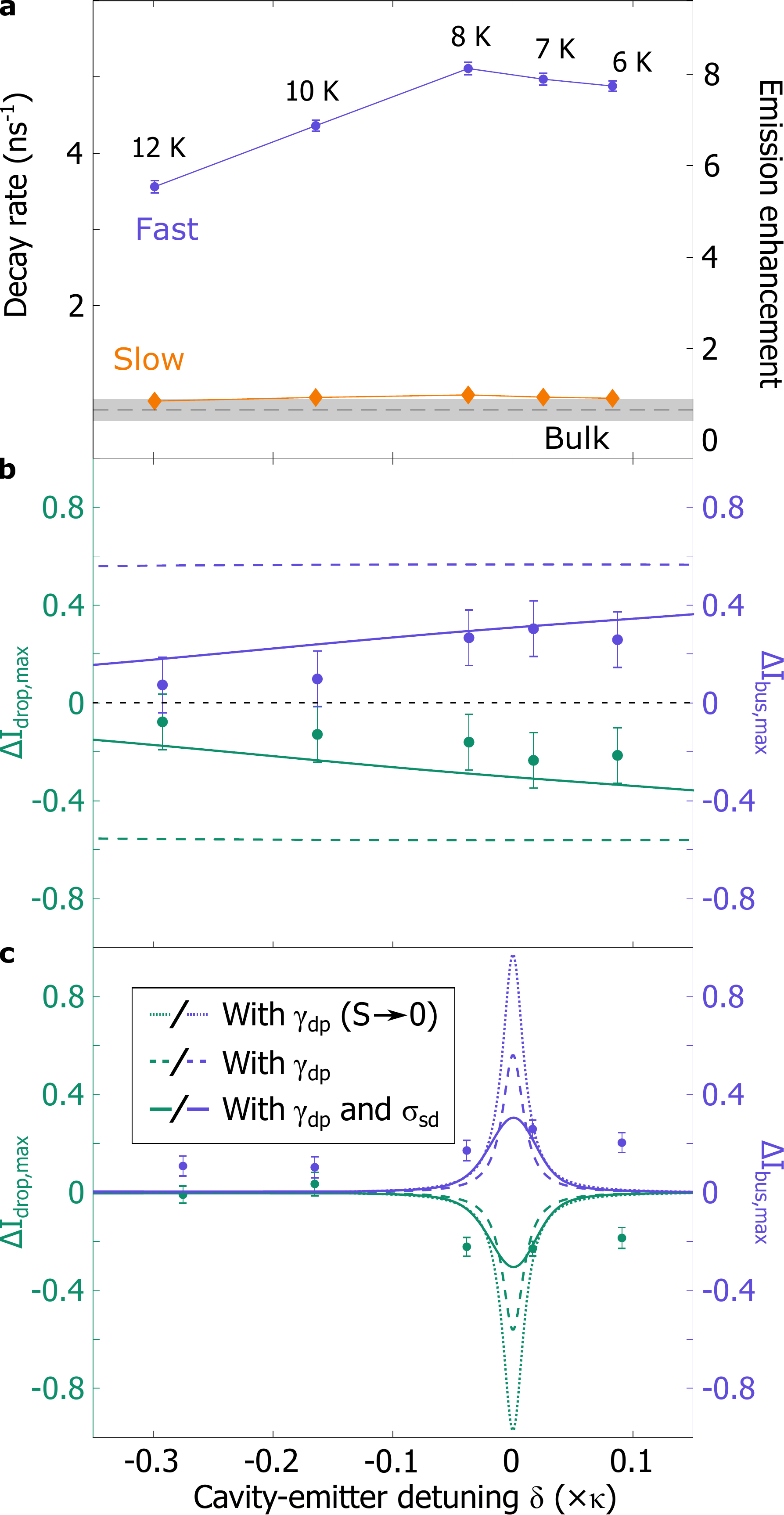}
  \end{center}
    \caption{(a) The QD lifetime as a function of QD-cavity detuning (in units of $\kappa$), with both the decay rate (left) and emission enhancement (right). The QD is temperature-tuned to the cavity resonance at 8K. Decay rates of both the well- (fast, purple) and weakly-coupled (slow, orange) dipoles are shown, and are compared to bulk emission rates (dashed line, shaded region depicts variance). Extinction as the QD is tuned across the cavity resonance, showing the maximum extinction $\Delta I_{\mathrm{drop, max}}$ at (b) the QD resonance and (c) the cavity resonance. }
  \label{lifetime}
\end{figure}

The photon routing can also be controlled by changing the QD-cavity detuning, which we do by varying the sample temperature, hence tuning the QD through the cavity resonance from -0.3$\kappa$ to 0.09$\kappa$ (Figure 4 in Supplementary Information). As the QD is scanned through the resonance, $\gamma_{\mathrm{fast}}$ increases as can be seen in Figure~\ref{lifetime}a (purple symbols) and peaks at a maximum lifetime of $\gamma_{\mathrm{fast}}$ = (5.11 $\pm$ 0.08) $\mathrm{ns}^{-1}$ at 8K ($\delta = -0.04\kappa$), corresponding to an 8-fold emission enhancement. In contrast, the weakly coupled transition decay rate (orange symbols) remains constant and near the bulk decay rate (shaded region).

We observe a similar trend in photon re-routing efficiency, shown in Figure~\ref{lifetime}b; as the QD becomes resonant with the microdisk, the maximum $\Delta I_{\mathrm{drop, max}}$ at the QD resonance ($\omega_{\mathrm{QD}}$) becomes increasingly negative (left axis, green symbols) while the maximum $\Delta I_{\mathrm{bus, max}}$ increases (right axis, purple symbols), as more photons are re-routed from the drop to bus port by the emitter, in good agreement with the theoretical calculations (solid curves). The predicted increase in routing efficiency for positive detunings is due to the decrease in spectral diffusion at lower temperatures (c.f. Supplementary Information). For our system, a maximum of $(23 \pm 3)\%$ of the photons are re-routed between the ports, although for a similar but electrically contacted resonator ($\sigma_{\mathrm{sd}}$ = 0 and $S = 1.5$, dashed curves), we predict that up to $56 \pm 3 \%$ of the photons can be re-routed (dashed curves).

Figure~\ref{lifetime}c demonstrates how our system can be used as a coherent photon router in practice, showing the fraction of photons scattered out of the drop port (left axis, green symbols) and into the bus port (right axis, purple symbols) for photons on resonance with the cavity ($\omega_{\mathrm{cav}}$) as a function of the QD-cavity detuning. Here, we observe that a detuning of the QD of $0.39\kappa$ (requiring a temperature change of only 6 K) is sufficient to completely turn off the router, corresponding to a shift of $5.1$ QD linewidths. For an electrically contacted resonator, the fraction of photons scattered into the bus port increases to $(56 \pm 3)\%$ (dashed curve), where the intensity of the incoming photon stream adds an additional control knob that increases this value to $ (97 \pm 2) \%$ (dotted curve). Instead of temperature tuning of the QD, it is also possible to achieve similar control electronically with a contacted sample \cite{Uppu:20} or even all-optically \cite{Turschmann:17}. 

\section{Discussion}
Enhancing the quantum light-matter interactions simultaneously increases the coupling of photons to desired modes and the coherence of the emission, with implications for a host of quantum technologies. We define this enhancement in terms of the Purcell factor $F$, which quantifies the change to the radiative emission rate, such that $\gamma_{\mathrm{cav}} = F \gamma_{\mathrm{bulk}}$. Experimentally, we find $F_{\mathrm{exp}}=6.9 \pm 0.9$ (c.f. Figure~\ref{lifetime}a), which can be compared to the predicted value of,\cite{Purcell:46,Wang:17}

\begin{equation}
  F_{\mathrm{ideal}} =\frac{3}{4\pi^2}\left(\frac{\lambda}{n}\right)^3\frac{Q_{\mathrm{exp}}}{V_{\mathrm{eff}}},
  \label{Purcell}
\end{equation}
where for our resonator $V_{\mathrm{eff}} \approx$ 18$(\lambda/n)^3$ and $Q_{\mathrm{exp}}$ = 8900 $\pm$ 100. The resulting $F_{\mathrm{ideal}}$ = 38 $\pm$ 1 is larger than measured experimentally due to spatial mismatch of the QD relative to the field maximum of the optical mode. Deterministic positioning\cite{Schnauber:18, Pregnolato:19} can address this issue.

\begin{figure}[t!]
  \begin{center}
    \includegraphics[trim={0 0cm 0 0},clip, width=0.45\textwidth]{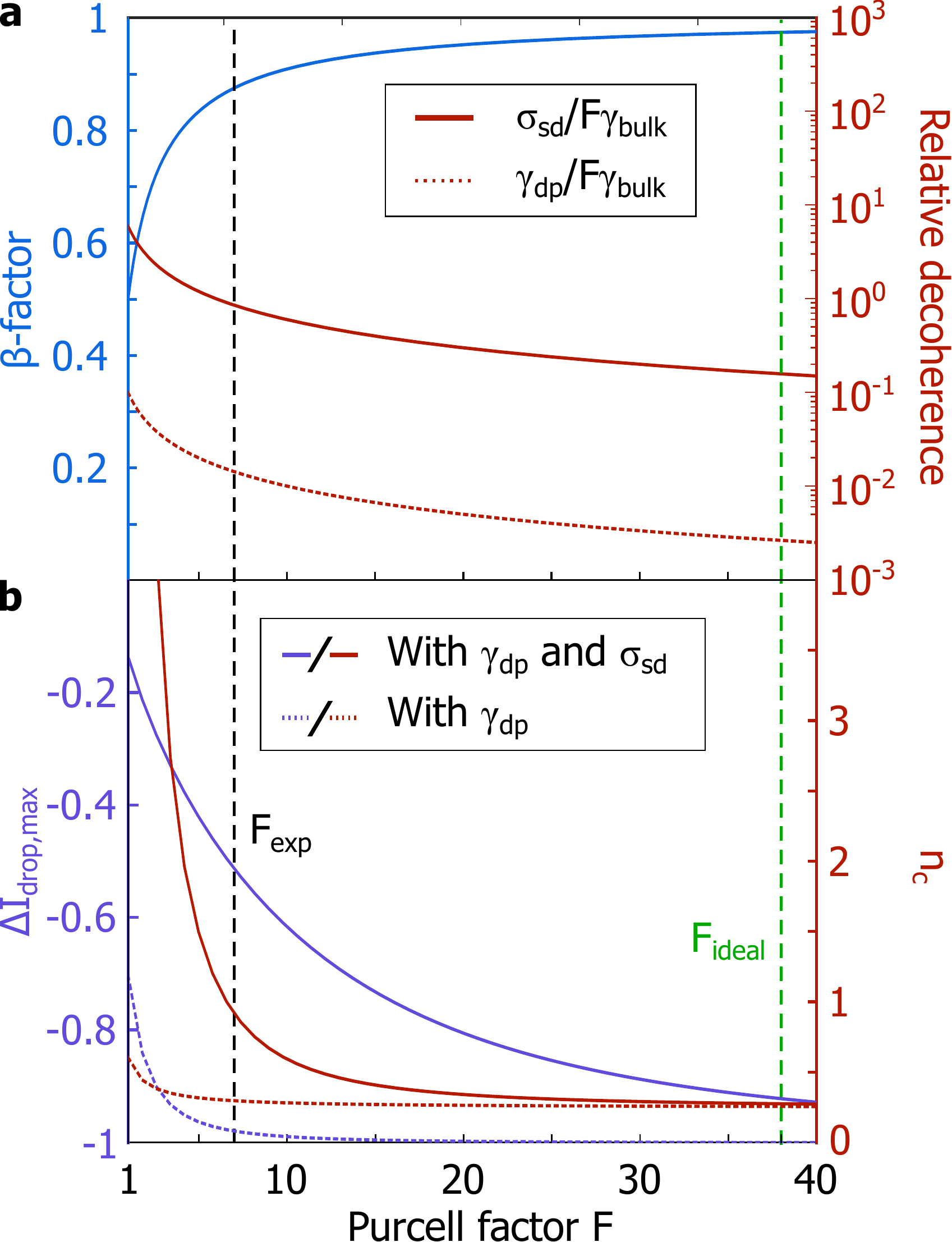}
  \end{center}
    \caption{Effect of the Purcell enhancement on system parameters. a) As $F$ increases, so too does the emitter-resonator coupling efficiency $\beta$ (left axis, solid). Conversely, an increase in the emission rate $\gamma_{\mathrm{cav}}= F \gamma_{\mathrm{bulk}}$ decreases the relative effect of the spectral diffusion $\sigma_{\mathrm{sd}}$ (right axis, solid) and dephasing $\gamma_{\mathrm{dp}}$ (right axis, dashed) to the system decoherence. b) The maximum drop port extinction (left) and critical photon number $n_c$ (right) achievable as a function of $F$, both with (solid) and without (dotted) spectral diffusion.}
  \label{beta}
\end{figure}

Increasing $F$ not only increases the emission rate, but also decreases the relative effect of decoherence mechanisms such as pure dephasing or spectral diffusion, with clear implications for bright sources of indistinguishable single photons \cite{Santori:02}. For our system, the dominant source of decoherence is spectral diffusion that, in bulk ($F$=1), results in a ratio of $\sigma_{\mathrm{sd}}/F \gamma_{\mathrm{bulk}} = $ 6. In Figure \ref{beta}a, we display how this ratio decreases as the $F$ increases, noting that moderate $F \geq $6 suffice to reach a unity ratio. A further two orders-of-magnitude reduction in decoherence is obtainable in the absence of spectral diffusion, motivating the use of electrically contacted resonators.

A QD-microdisk resonator system is not limited to acting as a source of quantum light states, but can also be used for their control or processing. As an example, this system can act as a Bell-state analyzer, a key element of a quantum optical network \cite{Lodahl:18}, either in a standard cavity-QED configuration \cite{Duan:04} or as a passive, nonlinear scatterer \cite{Ralph:15}. For cavity-QED, the Purcell factor can be re-expressed in terms of the QD-cavity coupling strength $g = \sqrt{F_{\mathrm{exp}}\kappa\gamma_{\mathrm{bulk}}}/2$,. The resulting $g/2\pi = 2.5 \pm 0.3$ GHz, which can be used to write the cooperativity of the system \cite{Borregaard:19} $C=4\left|g\right|^2 / \left[\kappa\gamma_{\mathrm{bulk}}\right] = 6.9 \pm 0.9$. Given the success rate of a cavity-QED based analyzer of $1 - 1/C$, we expect our modest $F_{\mathrm{exp}} = 6.9$ device to succeed $86\pm 11\%$ of the time.

On the other hand, the success of a Bell-state analyzer based on passive, coherent scattering from the QD depends on the emitter-waveguide coupling efficiency $\beta$ \cite{Ralph:15}. By expressing the (lower-bound) $\beta$-factor as
\begin{equation}
    \beta = \frac{\gamma_{\mathrm{cav}}}{ \gamma_{\mathrm{cav}}+ \gamma_{\mathrm{leak}}} = \frac{F}{F+1},
\end{equation}
we find an experimental $\beta_{\mathrm{exp}}\approx 0.87\pm 0.01$, which increases to $\beta_{\mathrm{ideal}}\approx 0.97$ for an optimally positioned QD. For this scheme, the success rate scales as $\left(2\beta - 1\right)/ \beta$, showing that near-perfect operation should be possible with our system.

Finally, as discussed in the main text, the QD-microdisk resonator can function as a coherent router, where the re-routing of photons between the drop and bus ports can be controlled either through the intensity of the incident photon stream or the QD-microdisk detuning. In Figure~\ref{beta}b we present the dependence of the maximum change in drop port intensity (i.e. re-routing efficiency, left axis) on the Purcell factor in the cases where the emitter suffers from both pure dephasing and spectral diffusion (solid curve) or just the former (dotted curve). Even for the non-contacted systems, we expect a re-routing efficiency in excess of $80\%$ for moderate enhancements of $F\approx20$, while for an electrically contacted sample near-perfect routing is predicted already at $F\approx 10$. We predict similar dependencies for the critical photon number (Figure~\ref{beta}b, right axis), where for an on-resonance emitter we observe that moderate Purcell factors are sufficient to overcome spectral diffusion for single-photon nonlinearities.

\section{Conclusions}
In summary, we present an integrated whispering gallery mode resonator system for on-chip quantum photonics based on single self-assembled QDs. For such a system, which can be easily integrated with other photonic components, Q-factors in excess of 20,000 are observed, enhancing emission into the desired optical modes to simultaneously achieve a high coupling efficiency $\beta> 0.85$ and compensate for the majority of the decoherence mechanisms. Using this platform, we demonstrate coherent re-routing of photons between the drop and bus ports, observing a peak efficiency of $(42\pm 4)\%$ that is expected to increase to $(53\pm 4)\%$ at $S=0$ and to $(98\pm 2)\%$ ($S=0$ and, $\sigma_{\mathrm{sd}}=0$) with electrical gating \cite{Uppu:20}. We show control over this routing using both temperature tuning and via the excitation intensity, with the latter requiring a critical photon number of only 0.94 photons per lifetime. Altogether, our platform enables coherent light-matter scattering \cite{lejeannic:21} and efficient quantum optical nonlinearities \cite{Javadi:15,Turschmann:19} at the single-photon level, two key functionalities of solid-state quantum technologies \cite{Borregaard:19}.


\begin{acknowledgement}
The authors gratefully acknowledge financial support from Danmarks Grundforskningsfond (DNRF
139, Hy-Q Center for Hybrid Quantum Networks) and the European Union's Horizon 2020 research and innovation programme under grant agreement No. 824140 (TOCHA, H2020-FETPROACT-01-2018). A.D.W. and A.L. acknowledge gratefully support of DFG-TRR160 and  BMBF - Q.Link.X  16KIS0867.
\end{acknowledgement}
\begin{suppinfo}
The supplementary information containing the theoretical models and fabrication details can be found here below.
\end{suppinfo}

\section{Supplementary Information}
\subsection{Raw bus port scans}

\begin{figure}
    \centering
    \includegraphics[trim={0 7.5cm 0cm 8.0cm},clip, width=0.8\textwidth]{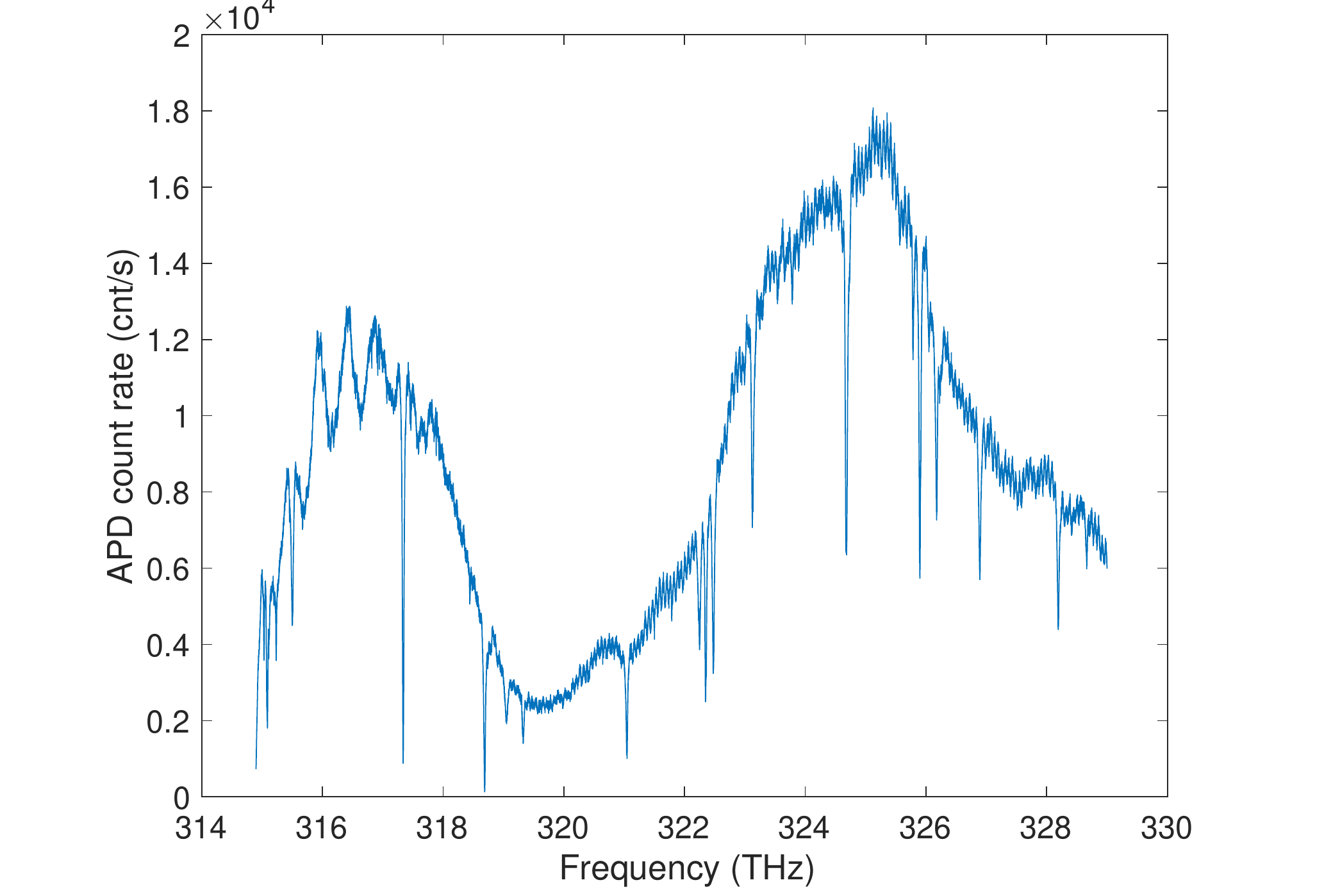}
    \caption{Exemplary transmission intensity as a function of frequency, using a resonant CTL laser and measuring the emission through the bus port. Transmission dips corresponding to cavity resonances are clearly visible, and their separations agree well with the calculated FSR of the disk.}
    \label{fig:busport_trans_raw}
\end{figure}

We use a tunable continuous (cw) laser (Toptica CTL) to study the sample by coupling the laser into the waveguide mode via the shallow etch gratings. We scan the laser frequency over a 13 GHz range and monitor the intensity that is outcoupled at the bus port with a single-photon detector. An example of the raw data is shown in Fig. \ref{fig:busport_trans_raw}, corresponding to Fig. 2a in the main manuscript. A large background oscillation due to the frequency dependent grating is visible. To make the transmission dips more comparable, they are each normalized to the background count rate at their frequency position, allowing us to obtain the dip transmission $\Delta T$. 

\subsection{Estimating the Purcell factor and coupling strength}

\begin{figure}
    \centering
    \includegraphics[trim={0 0 0 0},clip, width=0.5\textwidth]{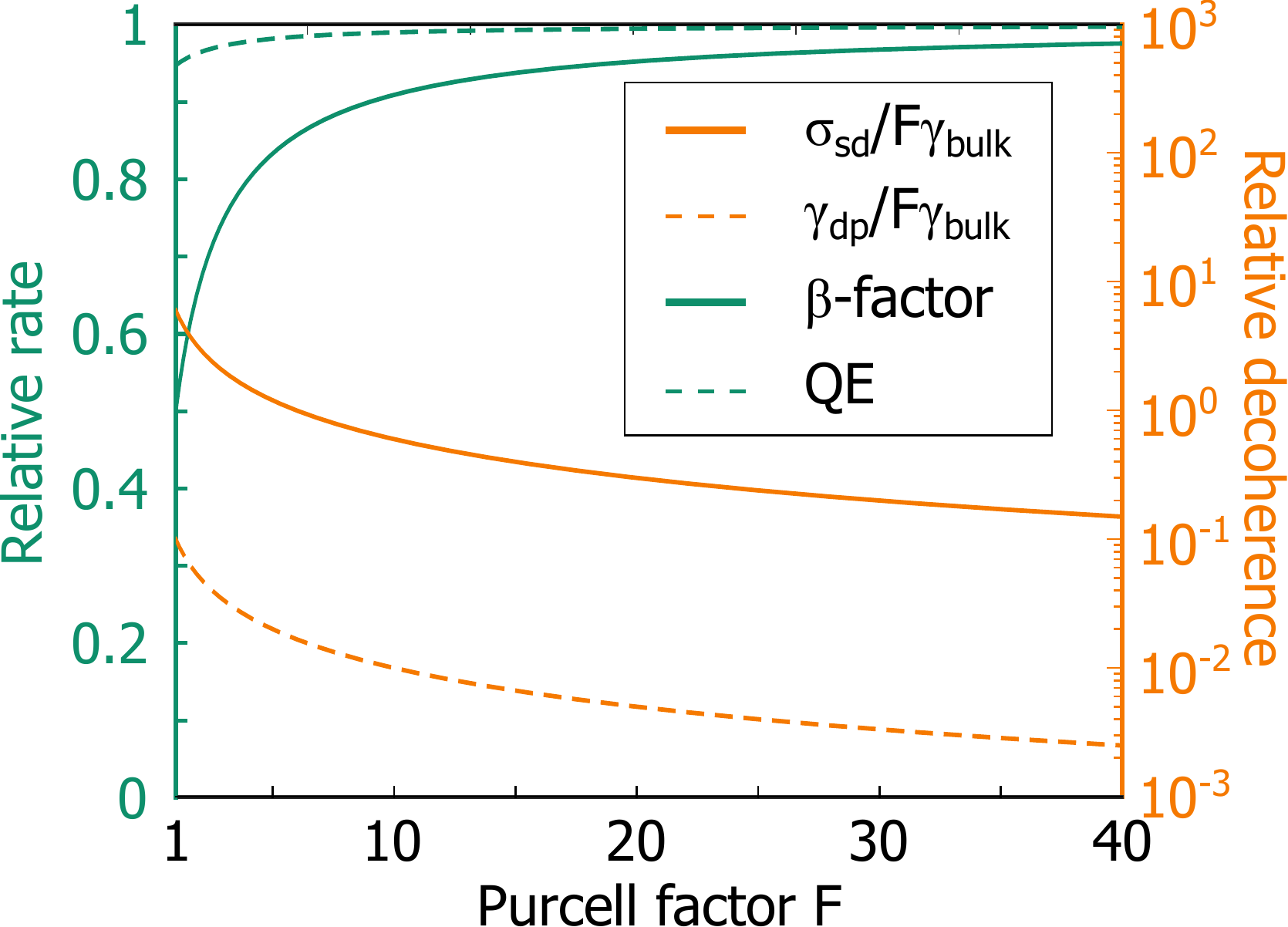}
    \caption{Effect of the Purcell enhancement on system parameters. As $F$ increases, so too does the emitter quantum efficiency (left axis, dashed) and the emitter-resonator coupling efficiency $\beta$ (left axis, solid). Conversely, an increase in the emission rate $\gamma_{\mathrm{cav}}= F \gamma_{\mathrm{bulk}}$ decreases the relative effect of the spectral diffusion $\sigma_{\mathrm{sd}}$ (right axis, solid) and dephasing $\gamma_{\mathrm{dp}}$ (right axis, dashed) to the system decoherence.}
    \label{fig:QE}
\end{figure}

The theoretical model in this paper is based on reference\cite{Garnier:07}. We start by considering a single Indium Arsenide (InAs) quantum dot (QD) in bulk Gallium Arsenide (GaAs). It decays with a rate of: $\gamma_\mathrm{bulk} = \gamma_0 + \gamma'$, corresponding to the radiative decay and decay to non-radiative channels respectively \cite{Johansen:08}. When the emitter is placed in a cavity, its overall decay rate is modified and is thus given by: 
\begin{equation}
    \gamma_\mathrm{tot} = \gamma_\mathrm{cav}+\gamma'+\gamma_\mathrm{leak}= F\gamma_0+\gamma'+\gamma_\mathrm{leak},
\end{equation}
where we consider decay into the cavity mode, non-radiative channels and non-cavity modes respectively. For InAs QDs in bulk GaAs, quantum efficiencies ($\mathrm{QE}_{\mathrm{bulk}}$) $>$0.9 are routinely reported \cite{Stobbe:09,Wang:11}. By coupling the QDs to a cavity, the enhanced radiative decay rate further increases its $\mathrm{QE}_{\mathrm{cavity}}$ as follows:
\begin{equation}
    \mathrm{QE}_{\mathrm{cavity}}=\frac{F\gamma_0+\gamma_\mathrm{leak}}{F\gamma_0+\gamma_\mathrm{leak}+\gamma'}=\frac{F\gamma_0+\gamma_0}{F\gamma_0+\gamma_0+(1-\mathrm{QE}_{\mathrm{bulk}})\gamma_\mathrm{bulk}}=\frac{\mathrm{QE}_{\mathrm{bulk}}(F+1)}{\mathrm{QE}_{\mathrm{bulk}} \cdot F+1},
\end{equation}

where we have used the approximation $\gamma_{\mathrm{leak}}=\gamma_{\mathrm{0}}$, as has been reported elsewhere \cite{Srinivasan:07}. The variation in $\mathrm{QE}_{\mathrm{cavity}}$ as a function of Purcell enhancement $F$ is displayed in Fig. \ref{fig:QE}, evaluated with $\mathrm{QE}_{\mathrm{bulk}}$=0.9 at $F=$1. Here the $\mathrm{QE}_{\mathrm{cavity}}$ increases above 0.99 for $F>$6.3, suggesting that the non-radiative decay rate is small and hence justifying the approximation $\gamma'\approx0$. 
By using these approximations, we can also write $\gamma_\mathrm{tot} \approx (F+1)\gamma_0$ and the coupling efficiency as \cite{Wang:17}:
\begin{equation}
    \beta=\frac{F\gamma_0}{\gamma_\mathrm{tot}}\approx\frac{F}{(F+1)}
\end{equation}
To obtain the Purcell factor experimentally, we measure the lifetime of the QD situated in the cavity and compare it to the average lifetime measured for QDs in bulk GaAs. A cavity-enhanced decay rate allows us to express:
\begin{equation}
\frac{\tau_\mathrm{bulk}}{\tau_\mathrm{tot}}=\frac{\gamma_\mathrm{cav}+\gamma'+\gamma_\mathrm{leak}}{\gamma_0 + \gamma'} \approx F+1
\end{equation}
Given the measured lifetimes, we obtain a $F_{\mathrm{exp}}=6.9 \pm 0.9$, with which we can further estimate the QD-cavity coupling strength as follows \cite{Wang:17}:
\begin{equation}
  g =\frac{\sqrt{F_{exp}\kappa\gamma_\mathrm{bulk}}}{2}
  \label{couplingrate}
\end{equation}
For our system, we obtain $\{g,\kappa,\gamma_\mathrm{bulk}\}/2\pi=\{2.5,36.6,0.1\}$ GHz. Subsequently, we estimate the Cooperativity factor $C$ using the following formula:
\begin{equation}
  C =\frac{4g^2}{\kappa\gamma_\mathrm{bulk}}\approx F_\mathrm{exp} = 6.9 \pm 0.9
  \label{Cooperativity}
\end{equation}

\subsection{Coherent interaction of the QD-cavity system}

\begin{figure}[t!]
 \includegraphics[trim={0 0cm 0 0},clip, width=1\textwidth]{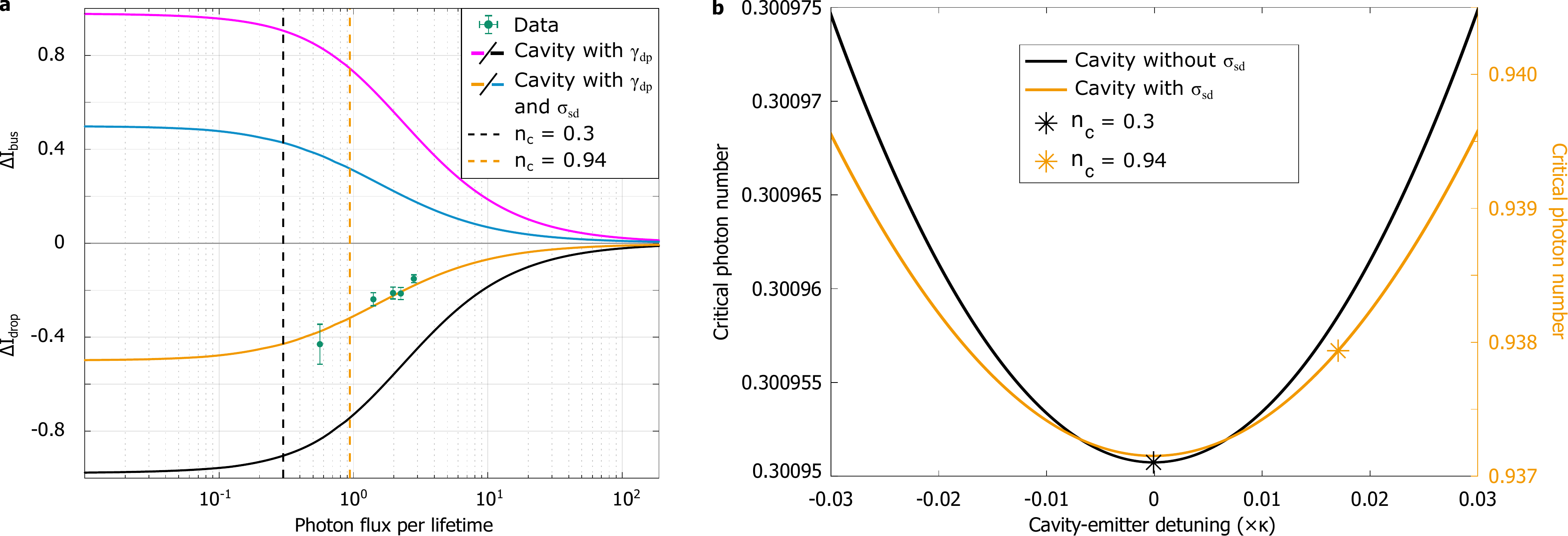}
  \caption{a) Power-dependent extinction of the QD of Fig. 3a in the main manuscript, showing a clear decrease in extinction for higher powers (green data points). Also shown are the theoretical $I_{\mathrm{drop}}$ and $I_{\mathrm{bus}}$ for the QD-cavity system accounting only for dephasing (black and pink respectively) and also for spectral diffusion (orange and blue). For the latter, a critical photon number of 0.94 photons per lifetime is found (orange dashed line). b) The variation in critical photon number $n_c$ as a function of QD-cavity detuning $\delta$. }
 \label{fig:sat}
\end{figure}

\begin{figure*}[t!]
 \includegraphics[trim={0 0cm 0 0},clip, width=1\textwidth]{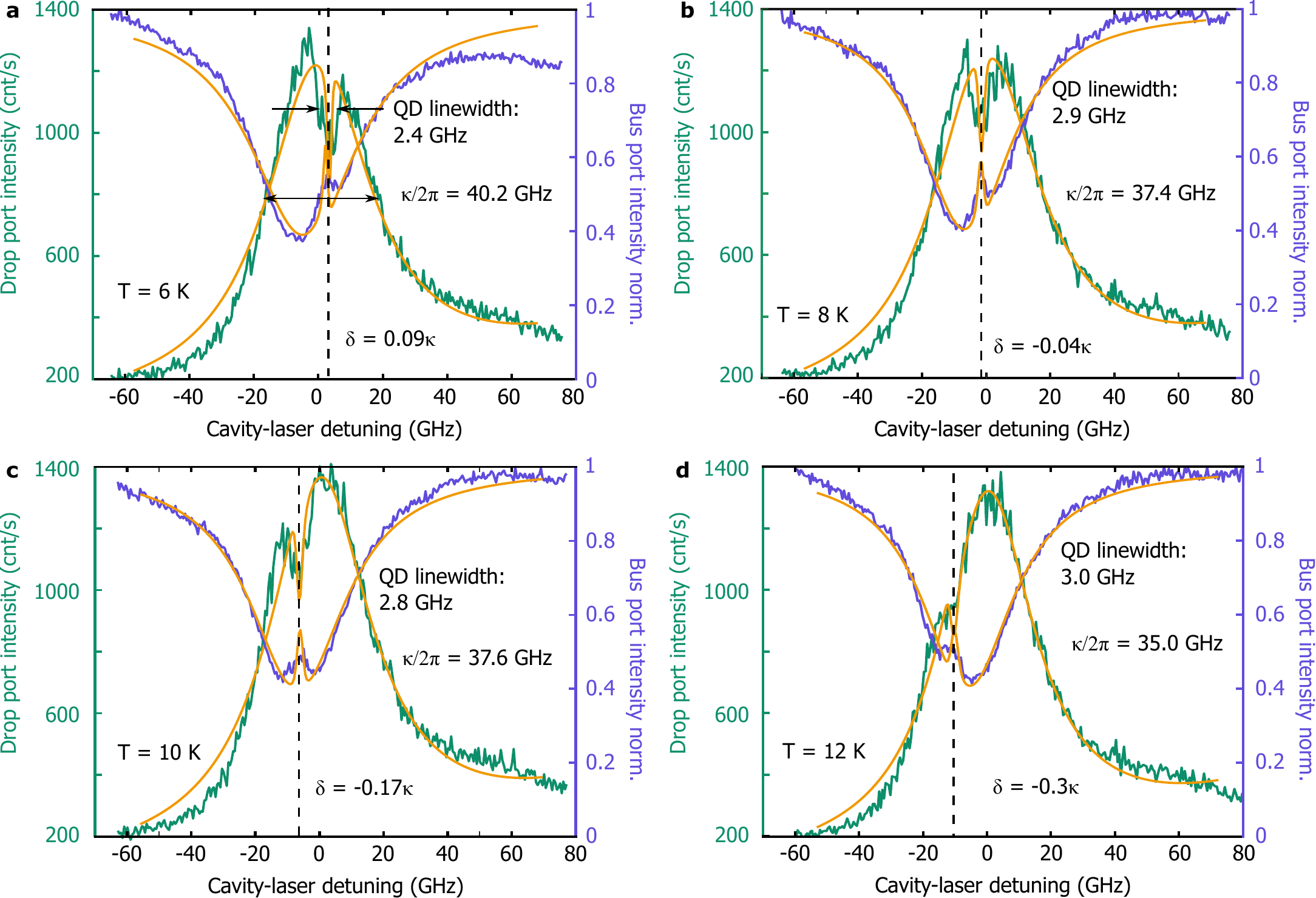}
  \caption{Temperature tuning of QD through the cavity resonance at a) 6 K, b) 8K, c) 10 K and d) 12 K. The solid curve is the theoretical fit. }
 \label{temperature}
\end{figure*}

To study the coherent interaction of the QD-cavity system, we consider a two-level system placed in a cavity and coupled to a waveguide, as depicted in Fig. 1 in the main text. We measure the transmission $T_{\mathrm{bus}}$ in the bus port and $T_{\mathrm{drop}}$ in the drop port. We start with the rate equations of the QD-cavity system, which are found in the literature \cite{Garnier:07}:
\begin{equation}
    \dot{s} = -i\Delta\omega s -\frac{\gamma_\mathrm{cav}}{2}\frac{Q}{Q_0}\left[t_0+\frac{1}{f} \right]s + i\frac{Q}{Q_0}\sqrt{\frac{\gamma_\mathrm{cav}}{2}}(2 s_z) b_{in}t_0
    \label{srate}
\end{equation}
\begin{equation}
    \dot{s_z} = -\gamma_\mathrm{cav}\frac{Q}{Q_0}\left[ \Re{(t_0)}+\frac{1}{f}\right]\left(s_z+\frac{1}{2}\right)+\sqrt{\frac{\gamma_\mathrm{cav}}{2}}\frac{Q}{Q_0}\left[i s^*b_{in}t_0+ c.c.\right]
    \label{szrate}
\end{equation}
\begin{equation}
    b_t = -\frac{Q}{Q_0}t_0b_{in}-i\frac{Q}{Q_0}\sqrt{\frac{\gamma_\mathrm{cav}}{2}}t_0 s,
\end{equation}
where $b_t$ is the outgoing field into the drop port, $b_\mathrm{in}$ is the incoming field amplitude, $Q$ is the total quality factor including coupling to leaky modes, $Q_0$ is the quality factor of the cavity mode, $t_0$ is the transmission of an empty cavity, $\Delta \omega = \omega_{laser}-\omega_{QD}$ is the frequency detuning of the drive field, $\kappa$ is the cavity linewidth, and $\delta$ is the detuning of the cavity resonance with respect to the QD resonance. The atomic operators are $S_z = \frac{1}{2}(\ket{e}\bra{e}-\ket{g}\bra{g})$ and $S_- = \ket{g}\bra{e}$ and in the above equations we are considering the expectation values, $s = \langle S_-\rangle$, $s_z = \langle S_z\rangle$, $b_t = \langle b_t\rangle$ and $b_{in} = \langle b_\mathrm{in}\rangle$. Here, $t_0$ is the bare cavity response in the absence of an emitter:
\begin{equation}
    t_0 = \frac{1}{1+i\frac{Q}{Q_0}\frac{\Delta\omega+\delta}{(\kappa/2)}}
\end{equation}
Additionally, the parameter $f$ describes the decay rate into the cavity versus all other rates:
\begin{equation}
      f = \frac{\gamma_\mathrm{cav}}{\gamma_\mathrm{leak}+2\gamma_\mathrm{dp}},\nonumber
\end{equation}
The steady-state solution to $s$ and $s_z$ can be found from Eqs.\ref{srate}-\ref{szrate} by setting $\dot{s}=0$ and $\dot{s_z}=0$, which results in:
\begin{equation}
    s = -\frac{2 i \sqrt{\frac{2}{\gamma_\mathrm{cav}}}s_z b_{in}t_0}{t_0 +\frac{1}{f}+\frac{2i\Delta\omega}{\gamma_\mathrm{cav}}\frac{Q_0}{Q}}
\end{equation}
\begin{equation}
    s_z = -\frac{1}{2}\frac{1}{1+\frac{|b_{in}|^2}{P_c}},
\end{equation}
\begin{equation}
    P_c = \frac{\gamma_\mathrm{cav}}{4|t_0|^2}\left[\frac{1}{f^2}+\frac{t_0+t_0^*}{f}+\frac{2i\Delta\omega}{\gamma_\mathrm{cav}}\frac{Q_0}{Q}(-t_0+t_0^*)+\left(\frac{2\Delta\omega}{\gamma_\mathrm{cav}}\frac{Q_0}{Q}\right)^2+|t_0|^2\right].
\end{equation}
where $P_c$ is the critical power to reach $s_z=-1/4$ and scales like the number of photons per second. Here we can use the following expressions to simplify $P_c$:
\begin{align}
    t_0+t_0^* &= 2\Re{(t_0)} \\
    t_0+t_0^* &= 2|t_0|^2 \\
    -t_0+t_0^* &= 2i\frac{Q}{Q_0}\frac{\Delta\omega+\delta}{(\kappa/2)}|t_0|^2 \\
    \frac{1}{|t_0|^2} &= 1+\left(\frac{Q}{Q_0}\frac{\Delta\omega+\delta}{(\kappa/2)}\right)^2
\end{align}
This allows us to obtain the following expression for the critical power $P_c$:
\begin{equation}
    P_c = \frac{\gamma_\mathrm{cav}}{4}\left[\left(1+\frac{1}{f}\right)^2+\left(\frac{Q}{Q_0}\frac{\Delta\omega+\delta}{f(\kappa/2)}\right)^2-\frac{4\Delta\omega}{\gamma_\mathrm{cav}}\frac{\Delta\omega+\delta}{(\kappa/2)}+\left(\frac{Q_0}{Q}\frac{2\Delta\omega}{\gamma_\mathrm{cav}}\right)^2+\left(\frac{2\Delta\omega}{\gamma_\mathrm{cav}}\frac{\Delta\omega+\delta}{(\kappa/2)}\right)^2\right].
    \label{FullPc}
\end{equation}
In the limit of $\gamma_\mathrm{leak},\gamma_\mathrm{dp}\rightarrow0$ and $Q_0=Q$, the above expression can be simplified to:
\begin{equation}
    P_c = \frac{\gamma_\mathrm{cav}}{4}\left[\left(\frac{2\Delta\omega}{\gamma_\mathrm{cav}}\right)^2+\left(\frac{2\Delta\omega}{\gamma_\mathrm{cav}}\frac{\Delta\omega+\delta}{(\kappa/2)}-1\right)^2\right]. 
\end{equation}
The atomic population $s$ therefore becomes:
\begin{equation}
   s = i\frac{Q}{Q_0}\sqrt{\frac{\gamma_\mathrm{cav}}{2}} b_{in} t_0 \frac{1}{1+S}\frac{1}{i\Delta\omega+ \frac{Q}{Q_0}\frac{\gamma_\mathrm{cav}}{2}(t_0+\frac{1}{f})},
   \label{population}
\end{equation}
where $S= \alpha |b_{in}|^2/P_c$ is the saturation parameter. We have also introduced a coupling efficiency $\alpha$ that relates the incoming light to the light that reaches the cavity, such that for an ideal lossless system $\alpha=1$. $S$ can hence be expressed as:
\begin{equation}
    S = \frac{\alpha n_{in}}{n_c},
    \label{Eq:S}
\end{equation}
where $n_{in} = |b_{in}|^2/\gamma_\mathrm{tot}$ and $n_c = P_c/\gamma_\mathrm{tot}$ are the number of incident photons per lifetime and the critical number of photons per lifetime to reach S = 1, respectively. In our work, we are considering a "leaky" cavity where the emitter in the cavity couples to leaky modes with decay rate $\gamma_\mathrm{leak}$ and experiences pure dephasing with the rate $\gamma_\mathrm{dp}$. Assuming coupling to the cavity via the waveguides only, we obtain the following transmission coefficient $t_\mathrm{drop} = b_t/b_\mathrm{in}$:
\begin{equation}
    t_\mathrm{drop} = \frac{b_t}{b_{in}} = t_0\left[-1+\frac{f}{\left(1+S\right)\left(f+\left(1+\frac{2i\Delta\omega}{\gamma_\mathrm{leak}+2\gamma_\mathrm{dp}}\right)\left(1+i\frac{\Delta\omega+\delta}{(\kappa/2)}\right)\right)}\right]
    \label{transcoeff}
\end{equation}

\begin{equation}
    t_\mathrm{bus}= 1+t_\mathrm{drop}
\end{equation}
This gives the steady-state cavity transmittivity in the drop and bus ports: $T_\mathrm{drop}=\chi|t_\mathrm{drop}|^2$ and $T_\mathrm{bus}=|t_\mathrm{bus}|^2$, where $\chi$ accounts for the total unnormalized count rate. Since the transmission by the QD depends on the incoming power, the first step in our fitting procedure is to use Eq. \ref{transcoeff} to fit the drop and bus port spectra at 7K whilst varying the power. We use the knowledge of the QD lifetime in the cavity and hence obtain the cavity linewidth $\kappa$, QD spectral diffusion $\sigma_{\mathrm{sd}}$, detuning $\delta$ and saturation parameter $S$. In Fig. \ref{fig:sat}a), we show the QD extinction for different powers together with its theoretical fit and observe experimentally that the coherent extinction by the QD on resonance with the cavity decreases as the power increases. Both the theoretical drop and bus port extinction $I_\mathrm{drop}$ and $I_\mathrm{bus}$, excluding spectral diffusion (pink and black solid lines), are displayed. Their inverse relation show that the incoming photons are either routed through to the bus port or drop port, and the ratio can be controlled by the incoming photon flux impinging on a single QD in the cavity, enabling its use as a photon switch. This analysis also allows us to obtain the critical photon number $n_c$ = 0.94 at a 7K at detuning $0.02\kappa$. In the absence of spectral diffusion and on resonance (but including dephasing), $n_c$ = 0.3, close to the ideal value of 0.25. 
Using the knowledge of $S$, we are further able to fit the temperature tuned data where the QD is moved through the cavity resonance, as displayed in Fig. \ref{temperature}.

For reference, the power-dependent transmission coefficient is further simplified when the emitter is resonant with the cavity, $\Delta\omega=\delta=0$:
\begin{equation}
    t_\mathrm{drop} = t_0\left[-1+\frac{f}{(1+f)(1+S)}\right].
    \label{satucurve}
\end{equation}

\subsection{Critical coupling}

The proximity of the waveguide to the cavity results in gap-dependent $\bar{Q}$, that increases as the gap between cavity and waveguide is widened. Following the formalism in reference \cite{Ding:10}, the coupling $\Delta T$ on resonance can be expressed as:
\begin{align}
    \Delta T &= 1-\left[T_{cc}+(1-T_{cc})\left(\frac{1-\kappa_g}{1+\kappa_g}\right)^2\right]\\
     \frac{1}{Q_{exp}} &= \frac{1}{Q_{int}}\left(1+\kappa_g\right)
    \label{criticalT}
\end{align}
where $\kappa_g = \kappa_{g0}e^{-\xi g}$ is the coupling rate between the cavity and access waveguides, $\xi$ is the characteristic length constant and $g$ is the gap size. In these expressions, $T_{cc}$ is the transmission at critical coupling, while $Q_{int}$ is the intrinsic quality factor of the resonator in the absence of the access waveguides.

\subsection{Background subtraction in data}
\begin{figure}[t!]
    \centering
    \includegraphics[scale=1]{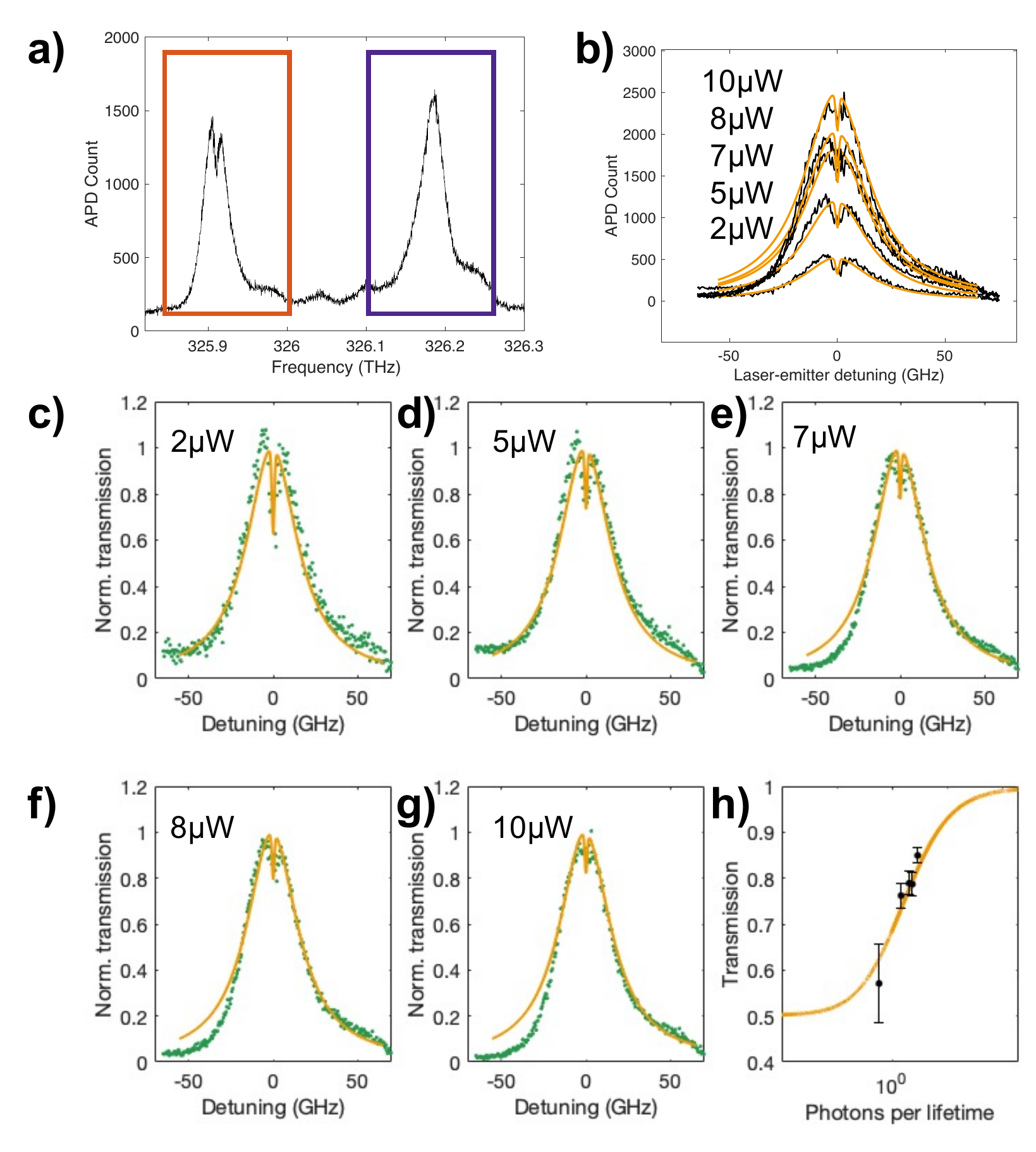}
    \caption{\textbf{a)} The spectra showing both the cavity resonance containing a QD (red box) and the neighbouring resonance (purple box). \textbf{b)} We fit the raw data by varying the power, including the existence of the second resonance. \textbf{c)-g)} The fitted data sets of different power, where second resonance has been subtracted and normalised. All 5 spectra are fitted together. \textbf{h)} The minimum data point for each power is plotted against saturation curve obtained from the power fit. }
    \label{fig:PowFitRout}
\end{figure}

The cavity resonance as shown in Fig. 3 in the main text is spectrally situated in the vicinity of another cavity mode, increasing the count rate on one side of the cavity mode, as depicted in Fig. \ref{fig:PowFitRout}a. To better fit our data, we include the second cavity mode in the fitting analysis, which is depicted in Fig. \ref{fig:PowFitRout}b where all power spectra are fitted simultaneously along with the second resonance. In order to model the power saturation curve given by Eq. \ref{transcoeff} at $S=0$, we subtract the additional counts from the second cavity mode using the double-cavity fit, which results in fits as shown in Fig. \ref{fig:PowFitRout}c-g. This data is subsequently fitted with Eq. \ref{transcoeff} convoluted with a Gaussian to include spectral diffusion, with which we also find the parameters $\alpha$, $\delta$, $\kappa$. Using Eq. \ref{FullPc} (assuming $Q=Q_0$) and Eq. \ref{Eq:S} we are able to convert input power to photons per lifetime and obtain Fig. \ref{fig:PowFitRout}h. Each data point in Fig. \ref{fig:PowFitRout}h is equal to the minimum  within the QD dip of the corresponding data. The errors are primarily due to the dark count noise on our single photon detectors.

\subsection{Temperature dependent spectral diffusion}
\begin{figure}
    \centering
    \includegraphics[scale=0.6]{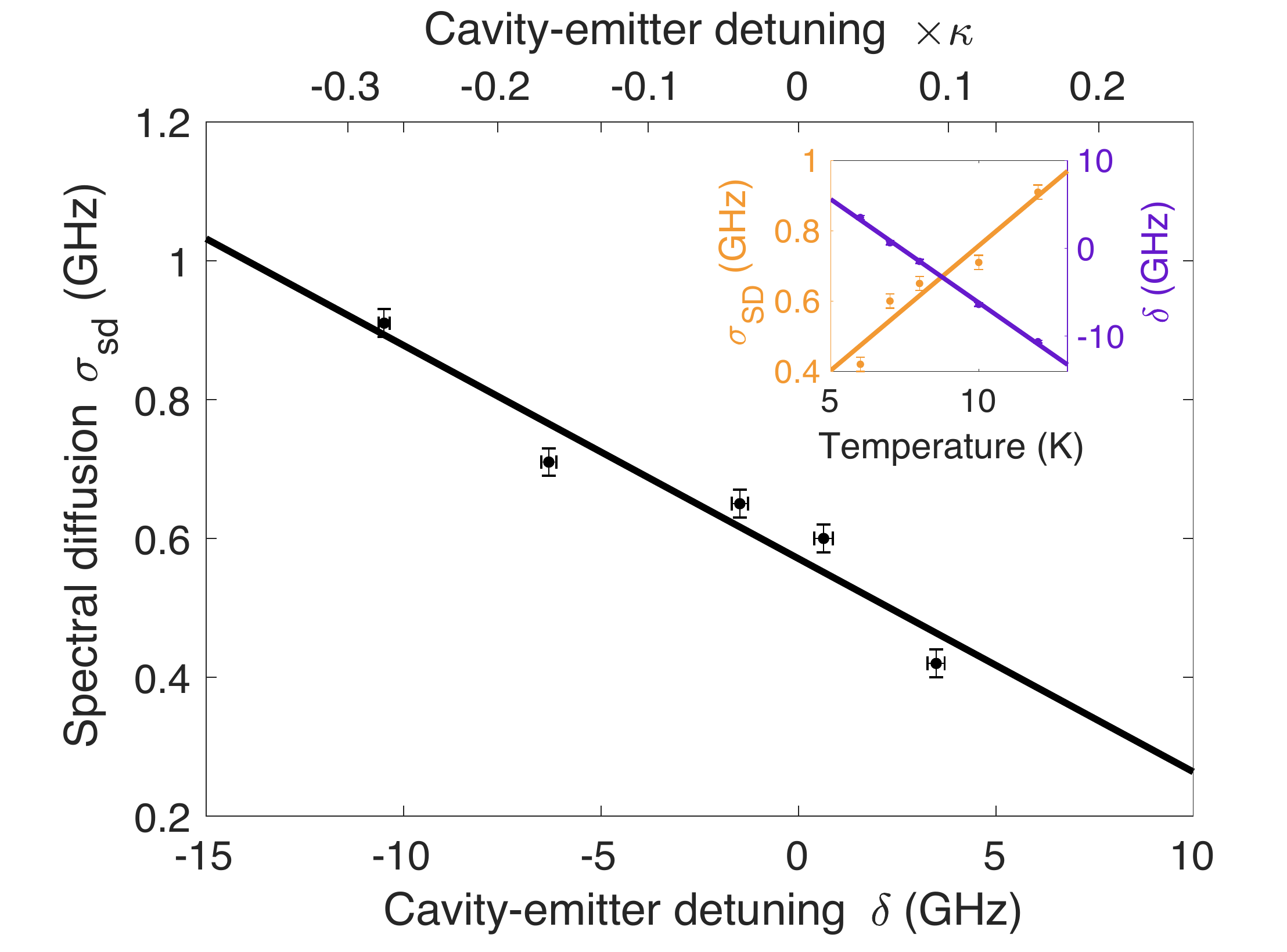}
    \caption{The amount of spectral diffusion depends on the cavity-QD detuning $\delta$ and follows a linear trend in this regime. Inset: This relation comes about due to temperature tuning, of which both the spectral diffusion $\sigma_{sd}$ and cavity-QD $\delta$ depend on.}
    \label{fig:TempSD}
\end{figure}

In our experiment, the QD is tuned across the cavity resonance by controlling the sample temperature and we denote the cavity-emitter detuning $\delta$. This is taken into account in the theoretical calculations displayed in Figure 4 in the main text, where the spectral diffusion $\sigma_{\mathrm{sd}}$ varies with temperature. From the inset in Fig. \ref{fig:TempSD}, it is shown that both $\delta$ and $\sigma_{\mathrm{sd}}$ scale linearly with temperature. This allows us to fit a linear relation between $\delta$ and $\sigma_{\mathrm{sd}}$, which we use to extract the amount of temperature-dependent spectral diffusion experienced by the QD.

\begin{figure}
    \centering
    \includegraphics[scale=0.8]{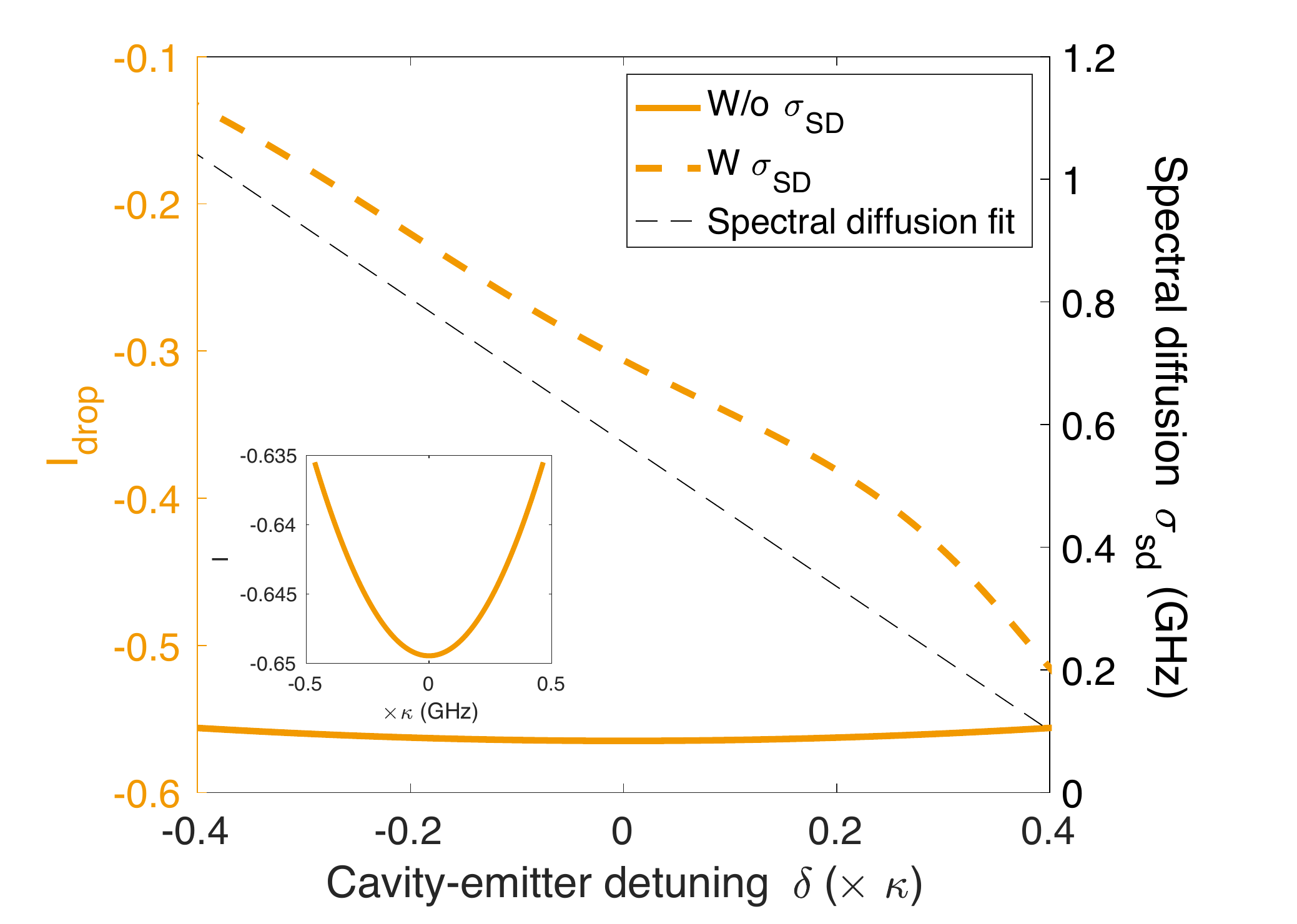}
    \caption{QD extinction $I_{\mathrm{drop}}$ with (orange dashed line) and without (orange solid line) spectral diffusion $\sigma_{sd}$. Since $\sigma_{sd}$ varies with temperature and is the main source of decoherence in our experiment, the maximum $I_{\mathrm{drop}}$ is not at cavity-emitter detuning $\delta = $ 0. Inset: Normalised QD extinction without spectral diffusion at saturation $S=1.5$ is expected to be maximized on resonance ($\delta = $ 0). The QD extinction is normalised to the bare cavity.}
    \label{fig:SanityExtinc}
\end{figure}

The linear relation of the spectral diffusion $\sigma_{sd}$ is taken into account in Fig. 4 in the main text, where the theoretical extinction of the QD is expected to be at maximum when the QD is on resonance with the cavity, as depicted on the inset in Fig. \ref{fig:SanityExtinc}. Due to the $\sigma_{sd}$ contribution, which decreases for lower temperatures, the extinction $I_{\mathrm{drop}}$ also scales linearly as can be seen of Fig. \ref{fig:SanityExtinc}.

\subsection{Mode profile and volume calculations}

A geometry representing the disc structures discussed in this paper was defined and meshed in COMSOL Multiphysics 5.1. Using the \textit{Electromagnetic Waves, Frequency Domain} and rotational symmetry we searched for eigenfrequencies in the disc in the 910-940nm range and were hence able to simulate the first and higher order modes of the structure. 

All variables are calculated using COMSOL along with the \textit{Electromagnetic Waves, Frequency Domain} physics package in order to obtain the final effective mode volumes of the various modes discussed in the main manuscript, using n = 3.46 for GaAs. In order to obtain a sufficiently small convergence error in the simulations, the geometry was meshed to a maximum element size of $36nm$ (corresponding to $\frac{1}{5}$ of the height of the disc), while the pedestal and the surrounding air was meshed to a maximum element size of $230nm$ (corresponding to $\frac{1}{4}$ of the central wavelength within the simulation). Finally, a perfectly matched layer (PML) enclosing the geometry and air is added as the outer boundary of the setup. This procedure results in a convergence error $<10^{-7}$. 

We follow the approach presented in \cite{Martincano:19} but adapt it for a linear dipole, using the knowledge of how the counter propagating modes are related, and given the mode volume when the dipole is placed in the field maximum, we obtain a minimum mode volume for a lossy structure in cylindrical coordinates:
\begin{equation}
    V=\frac{\pi\int\int drdz\cdot r\cdot [\epsilon(-E_r^2-E_z^2+E_{\phi}^2)-\mu(H_r^2+H_z^2-H_{\phi}^2)]}{2\epsilon_0n^2[(\mathrm{max}(E_r)]^2},
\end{equation}
where $E$ is the electric field while $H$ is the magnetic field of the mode. The factor $n$ is the refractive index, $\epsilon$ is the permittivity of the material while $\epsilon_0$ is the vacuum permittivity, and $\mu$ is the permeability of the material. 

\begin{figure}
    \centering
    \includegraphics[scale=0.5]{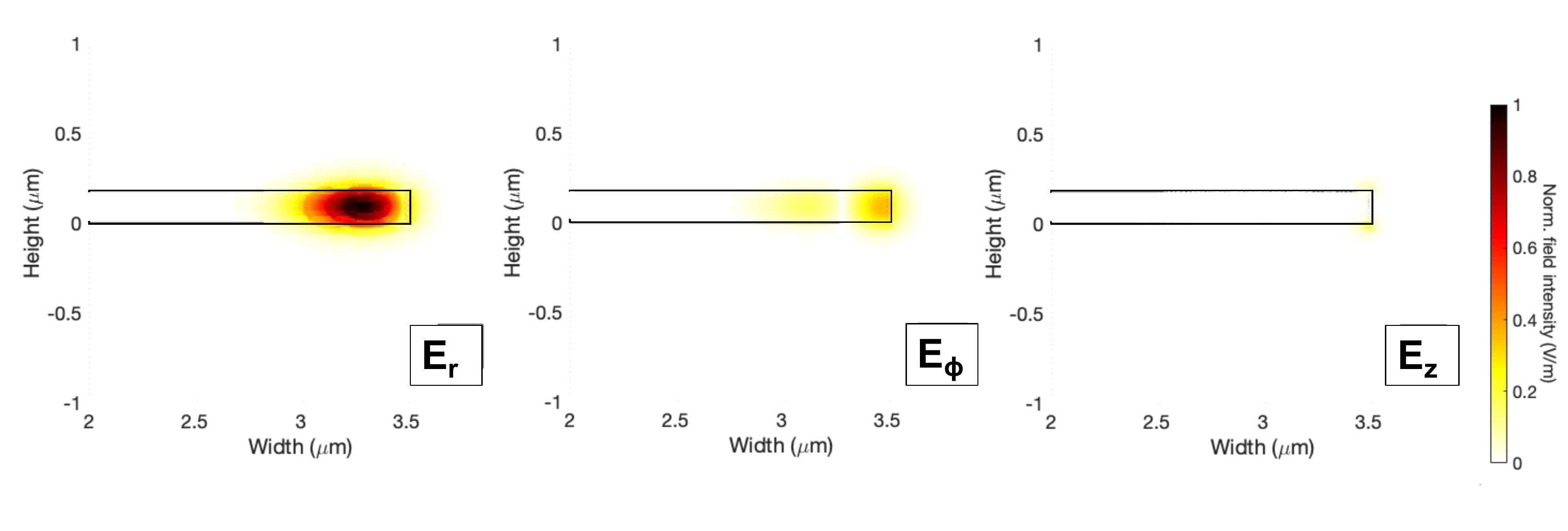}
    \caption{Spatial components of the first order electric field normalized to the absolute maximal field strength.}
    \label{fig:COMSOLModes}
\end{figure}

The normalized mode profile is presented in the main text. In Fig. \ref{fig:COMSOLModes} we see the contributions from the various components of the mode and note that the radial component is by far the dominant mode as would be expected.

\subsection{Fabrication}
\begin{figure}
    \centering
    \includegraphics[scale=0.63]{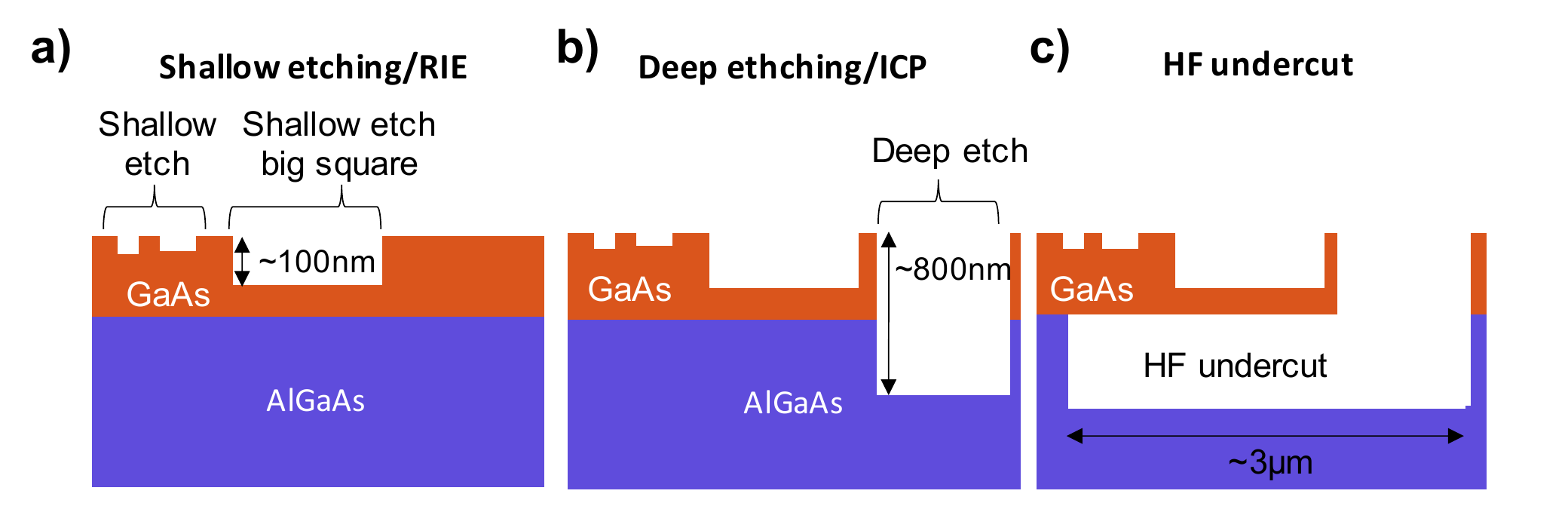}
    \caption{Three main steps in fabrication of microdisc resonators. a) Shallow etching a large square to a target depth of around 100nm determined by reflectivity measurements done during the RIE process. The small pattern of the gratings obtain a final etch depth around 50nm-60nm. b) Another layer is etched using a deep ICP etch to a target depth at around 800nm, well within the $AlGaAs$ layer. c) A solution of 10\% HF acid allows us to underetch through the deep trenches. An underetching of 50s seconds produces approximately a $3\mu m$ long undercut.}
    \label{fig:FabProcess}
\end{figure}

Disc resonators used throughout this study were fabricated on an undoped GaAs (160nm)/ AlGaAs (1150nm) wafer embedded with InAs quantum dots. The devices were developed using Electron Beam Lithography (EBL) on a layer of resist (ZEP520). Our smallest design features are 40 nm, which is larger than the precision of the EBL-alignment error of $\approx$ 30 nm. The shallow edged gratings were etched using Reactive Ion Etching (RIE), followed by an Inductively Coupled Plasma (ICP) etching to a depth of 800 nm. In the final step, the remaining structures were under-etched to fully suspend the waveguides and the periphery of the discs ($\approx 3\mu$m) using hydrofluoric acid (HF, 10\%), which is depicted in Fig \ref{fig:FabProcess}. The under-etching step sets a lower limit on the size of the disc. Following the under-etching, the sample was dried using a Critical Point Dryer (CPD).

\bibliography{bib_integrated_whispering-gallery-mode_resonator_for_solid-state_coherent_quantum_photonics}
\end{document}